# The Piezochiral Effect

Z. Zeng[1,2], M. Först[1], M. Fechner[1], X. Deng[1,2], A. Cavalleri[1,2], P. Radaelli[2]

[1]*Max Planck Institute for the Structure and Dynamics of Matter, Hamburg, Germany*
[2]*Department of Physics, Clarendon Laboratory, University of Oxford, Oxford, United Kingdom*

**Chirality is a pervasive property of matter that underpins many important phenomena across physics[1], chemistry[2] and biology[3]. Given its broad significance, the development of protocols for rational control of chirality in solid state systems is highly desirable, especially if this effect can be tuned continuously and in two directions. Yet, this goal has remained elusive due to the absence of a universal conjugate field that couples linearly to this structural order[4,5]. Here, we introduce the piezochiral effect, which enables control of chirality through mechanical strain. We first show by symmetry analysis that uniaxial strain induces chirality in a broad class of achiral crystals that host fragments of opposite chirality within each unit cell[6,7], an effect that has so far remained unrecognized. The strain-induced handedness can be tuned either by changing the strain direction or by switching between compressive and tensile strain. We experimentally verify this effect in $AgGaS_2$, using measurements of the optical activity under strain. Our discovery establishes a new scheme for chirality control, with potential applications that range from spintronics to asymmetric catalysis, and enantioselective interactions in biosystems.**

Chirality, the geometric property of an object that cannot be superimposed onto its mirror image[8], plays a central role across the physical and life sciences. It forms the basis of multiple functionalities such as magneto-chiral anisotropy[1], asymmetric synthesis[2] and enantioselective drug action[3]. Given its importance, extensive efforts have been devoted to chiral material engineering, primarily through bottom-up approaches, including the crystal growth[9], molecular synthesis[2] and nanostructure fabrication[10]. By comparison, the alternative route of inducing and controlling chirality in achiral systems by external fields has remained far less explored[11-13]. A key challenge lies in the absence of a universal chiral field that couples directly to handedness[4,5], unlike other functionalities such as electric polarization and magnetization, which can be readily controlled by electric and magnetic fields, respectively. Here, we uncover a scheme for chirality control by introducing the piezochiral effect, which describes the linear coupling between mechanical strain and chirality.

**Material classes with a piezochiral effect**

Strain engineering has long been recognized as a powerful means to tailor material functionalities[14-17]. Well-established examples, such as piezoelectricity[18] and piezomagnetism[19], have been extensively studied and led to widespread applications[20]. However, the linear coupling of strain to chirality has not yet been investigated. Whereas strain, polarization, and magnetization are directional quantities, chirality is a non-directional pseudoscalar that reverses sign under inversion symmetry[21]. This distinction renders the coupling between strain and chirality elusive.

The piezochiral effect is defined here as the induction of chirality in an achiral system by the application of strain (Fig. 1a). Mathematically, the effect can be written as

$$C = \sum_{ij} T_{ij} \varepsilon_{ij} \quad (1)$$

where $C$ denotes the strain-induced chirality, $\varepsilon_{ij}$ is the strain tensor and $T_{ij}$ is the piezochiral tensor, which we introduced here to describe the coupling between strain and chirality (See Supplementary Information). The sign of the strain tensor reverses upon switching between compressive and tensile strain, allowing induction of left and right handedness on demand, analogous to the reversal of the electric polarization and magnetization in piezoelectricity and piezomagnetism, respectively.

Building on this formalism, we performed a systematic group theory analysis. We enumerated all the achiral crystallographic point groups and identified the cases where the piezochiral effect is symmetry-allowed (Fig. 1b). For each allowed group, we derived the explicit form of the piezochiral coupling consistent with the symmetry constraints. For instance, in point group $\bar{4}2m$ (D2d), the piezochiral coupling takes the form

$$C = \theta(\varepsilon_{xx} - \varepsilon_{yy}) \qquad (2)$$

with $\theta$ the material-specific coupling coefficient. This coupling relation suggests that chirality can be induced by applying uniaxial strain along either the $x$ or $y$ direction in this point group. Since $\varepsilon_{xx}$ and $\varepsilon_{yy}$ enter the coupling with opposite signs, applying the same type of strain (compressive or tensile) along the two directions induces chirality of opposite handedness.

The framework of the piezochiral effect generalizes to higher-order couplings beyond the linear coupling, where chirality couples to quadratic or even higher-order strain terms. This behavior is directly analogous to higher-order piezoelectricity[22] and higher-order piezomagnetism[23], and becomes crucial in materials where the linear term is symmetry-forbidden.

The general form of the higher-order couplings can be expressed as

$$C = \sum_{ij} T_{ij}\varepsilon_{ij} + \sum_{ijkl} Q_{ijkl}\varepsilon_{ij}\varepsilon_{kl} + \cdots \qquad (3)$$

Through an analysis of all achiral crystallographic point groups, we identify four groups in which second-order piezochiral coupling provides the leading contribution (Table S3 in Supplementary Information), including materials like GaSe (point group -6m2). In addition, two groups allow third-order piezochiral coupling as the leading term (Table S4 in Supplementary Information), including materials such as GaAs (point group -43m).

**Experimental verification in the model system AgGaS$_2$**

We investigate the model system AgGaS$_2$ (point group $\overline{4}2m$) to uncover the microscopic origin of the piezochiral effect at the unit-cell level. The crystal structure of AgGaS$_2$ contains chiral substructures of opposite handedness within each unit cell, shown in red and blue in Figure 2a. These substructures adopt an antiferrochiral arrangement, in which the opposite handedness compensates, resulting in an overall achiral structure[6,7].

Next, we define a geometrical order parameter to quantitatively describe the chirality in the crystal. We choose a pseudoscalar defined as the triple product

$$C = (\widehat{r_1} \times \widehat{r_2}) \cdot \widehat{r_3} \qquad (4)$$

of the normalized vectors connecting the four neighboring atoms in a spiral structure[24] (Fig. 2b). This order parameter serves as an effective measure of chirality, vanishing in an achiral structure, changing sign for left- and right-handed enantiomers and remaining invariant under rotations. In AgGaS$_2$, the order parameter is evaluated for each substructure by summing over the four neighboring atoms in the spiral. The resulting values are opposite for the substructures of opposite handedness. The sum of these substructures yields zero net chirality, indicating an achiral unit cell (Fig. 2c).

We employed density functional theory calculations to examine the emergence of chirality in AgGaS$_2$ under uniaxial strain by evaluating the chiral order parameter after full structural relaxation of the strained crystal (see Supplementary Information). The calculations reveal that uniaxial strain along the $x$ direction modulates the chirality amplitude in both the opposite-handedness substructures (Fig. 3a). Importantly, the degeneracy between the left- and right-handed substructures is lifted, yielding a finite net chirality of the unit cell, with its handedness determined by the sign of the applied strain. The lifted degeneracy is reflected in a view of the AgGaS$_2$ crystal along its $c$ axis, which clearly shows the red and blue substructures distorted into different shapes (Fig. 3a inset). When strain is applied along the orthogonal $y$ direction, the sign of the induced chirality is reversed (Fig. 3b), consistent with the symmetry-derived form of the piezochiral coupling discussed above.

We experimentally verified the piezochiral effect by measuring the optical activity of a $c$-cut AgGaS$_2$ single crystal under strain, which directly reflects its geometrical chirality[25,26]. The polarization rotation of near-infrared light, measured as a function of the incident probe polarization, enabled us to isolate the signal contribution of the optical activity from birefringence effects, which are also induced by the application of strain (Fig. 4a). Figure 4b shows the results for uniaxial strain applied along the $x$ direction. The optical activity is extracted from the data by averaging the polarization rotation signal over all the incident probe polarizations, which is possible because the strain-induced birefringence signal averages to zero[27,28] (see Supplementary Information).

The extracted optical activity shows a linear dependence on the applied strain, providing direct evidence of the piezochiral effect (Fig. 4c). Remarkably, the handedness of the chiral state is reversed between the compressive and tensile strain applied, highlighting the versatility of the effect in inducing chirality. Further experiments with strain applied along the orthogonal $y$ direction show a sign change in the induced optical activity compared to $x$-axis strain (Fig. 4d & 4e). Quantitatively, a uniaxial strain of only 0.02% yields a rotary power on the order of 10% of α-quartz[28], a widely used chiral optical material, demonstrating the large magnitude of the piezochiral response in $AgGaS_2$. In combination, these experimental results provide clear verification of strain-induced chirality in $AgGaS_2$, with the symmetry in agreement with the theoretically predicted piezochiral effect.

**Summary and outlook**

Starting from symmetry considerations, we have introduced the piezochiral effect and established a new formalism to describe the linear coupling between strain and chirality. This effect was verified in the prototype compound $AgGaS_2$ through measurements of the strain-induced optical activity. Our discovery expands the family of strain-responsive functionalities by introducing a new member alongside piezoelectricity[18] and piezomagnetism[19], thereby opening new possibilities across a broad spectrum of applications. Importantly, chirality can be induced by uniform strain within this framework, in contrast to torsional strain, which involves a strain gradient that breaks translational symmetry[29].
One particularly exciting direction involves the integration of piezochiral materials into microelectromechanical systems (MEMS)[30], enabling chirality control within compact on-chip platforms. Such integration could lead to the development of enantioselective biosensors[31], dynamically tunable optical devices and polarization-based information technologies.

Asymmetric catalysis represents another compelling avenue for the piezochiral effect[32,33]. One strategy is to employ piezochiral materials directly as asymmetric catalysts, where strain-controlled chirality could enable rational control over enantioselectivity in specific reactions. Another strategy is to use piezochiral substrates as scaffolds for well-established asymmetric catalysts[34], providing a means to dynamically tune enantiomeric excess and selectivity, thereby addressing long-standing goals of controllability and versatility in asymmetric synthesis[35].

Beyond technological applications, the piezochiral effect could serve as a unique channel for coupling strain to emergent chiral order in quantum materials[36-40], opening opportunities to probe and manipulate novel quantum phases. More broadly, the piezochiral effect offers a new framework for chiral matter engineering, with interdisciplinary implications for photonics[41], spintronics[42,43], biosensing[44], and quantum information[45].

**Figures:**

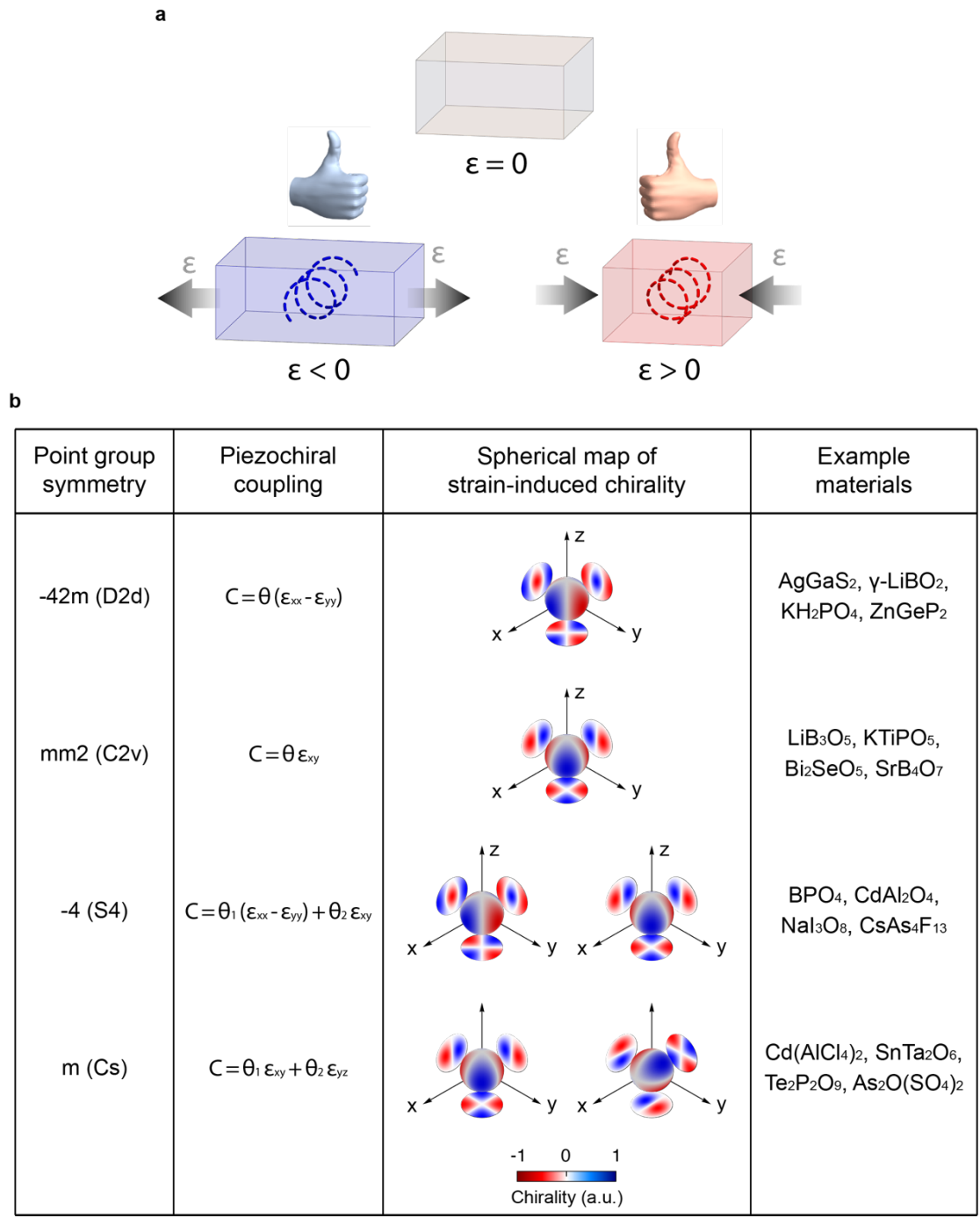

**Fig. 1 | The piezochiral effect. a,** An achiral system at equilibrium develops chirality under strain. Switching between tensile strain and compressive strain reverses the handedness of the induced chirality. **b,** Piezochiral coupling forms. From left to right: four achiral point groups supporting coupling between strain and chirality, their respective coupling forms, visualization of the coupling forms, and representative materials.

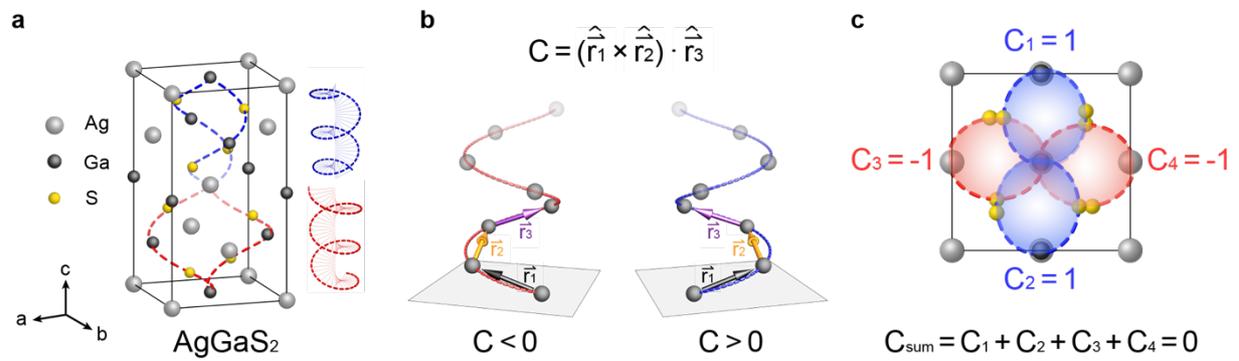

**Fig. 2 | Static chiral order parameter in AgGaS$_2$. a,** Unit cell of the antiferrochiral crystal AgGaS$_2$, composed of chiral substructures with opposite handedness. **b,** Definition of the chiral order parameter as the triple product of the normalized vectors between the neighboring atoms along a spiral. **c,** Normalized chiral order parameter of AgGaS$_2$. Substructures of opposite handedness yield opposite values, resulting in a vanishing net chiral order parameter.

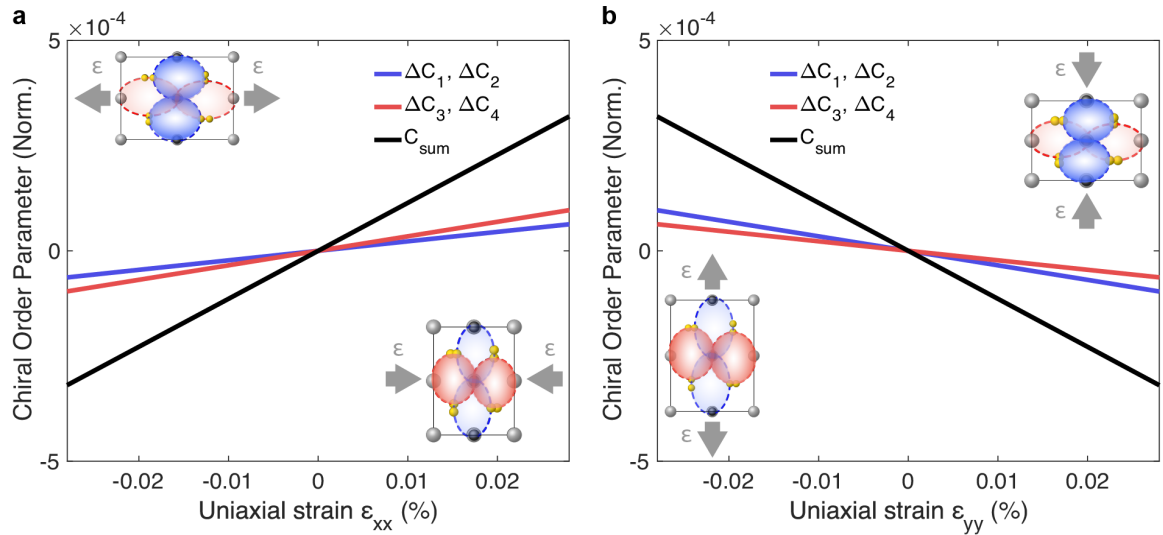

**Fig. 3 | Calculation of strain-induced chirality in AgGaS$_2$. a,** Evolution of chirality in the chiral substructures and in the overall unit cell under uniaxial strain along the x direction. **b,** Same as (a), but under uniaxial strain along the y direction.

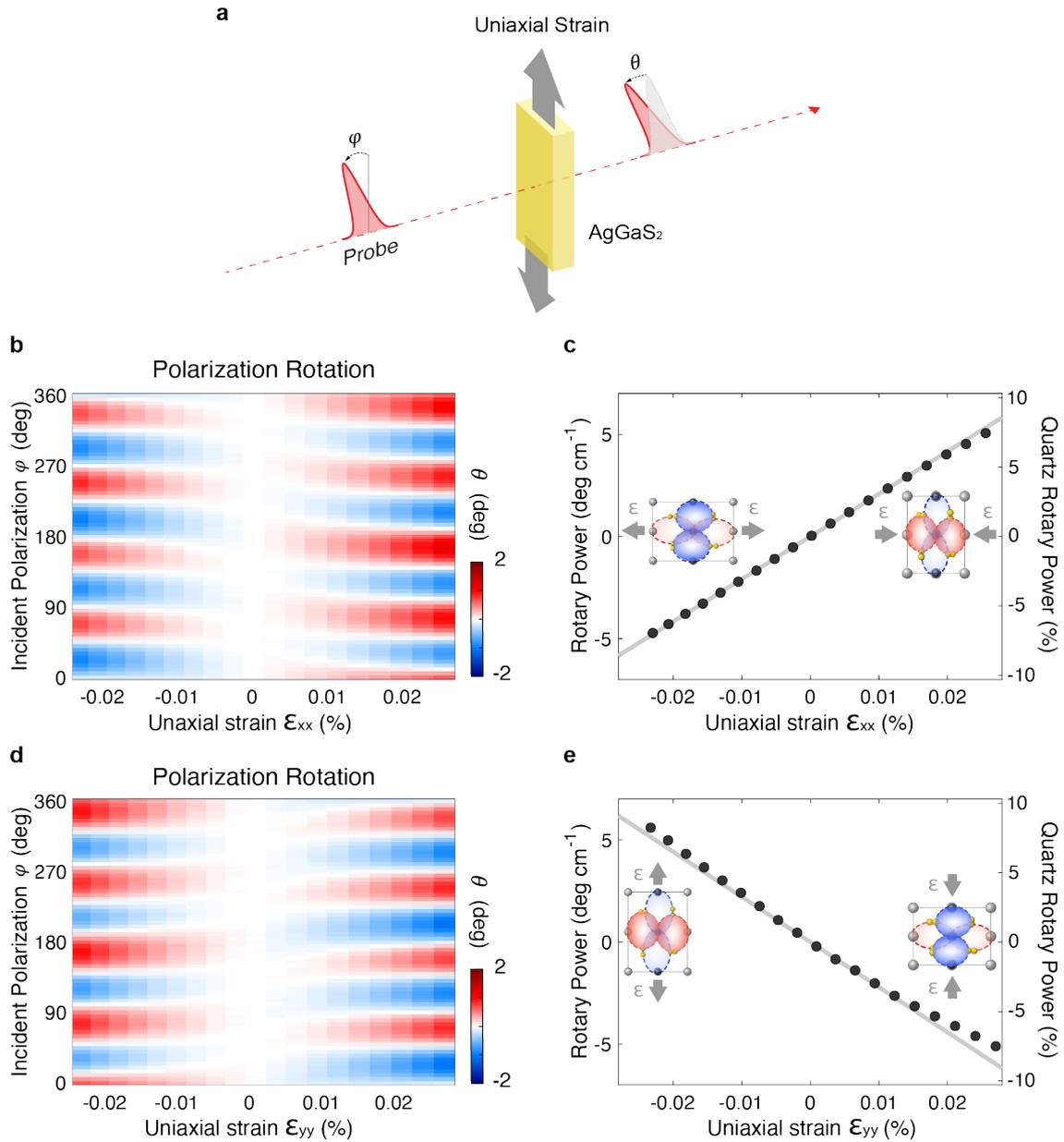

**Fig. 4 | Observation of piezochiral effects. a,** Schematic of the experiment. A linearly polarized optical pulse, propagating along the crystal c axis, probes strained-induced chirality by polarization rotation, measured as a function of the probe incident polarization. **b,** Polarization rotation signal for uniaxial strain along the x direction as a function of probe incident polarization. **c,** Rotary power as a function of uniaxial strain, extracted from (b). **d,** Same as (b), but for uniaxial strain along the y direction. **e,** Same as (c), but for uniaxial strain along the y direction.

# Supplementary Materials for

# The Piezochiral Effect

Z. Zeng[1,2], M. Först[1], M. Fechner[1], X. Deng[1,2], A. Cavalleri[1,2], P. G. Radaelli[2]

[1]*Max Planck Institute for the Structure and Dynamics of Matter, Hamburg, Germany*
[2]*Department of Physics, Clarendon Laboratory, University of Oxford, Oxford, United Kingdom*

**This file includes:**

Sections S1 to S4

Figs. S1 to S4

Tables S1 to S4

References

## S1. Materials and methods

### S1.1. Experimental Setup

Figure S1 shows the schematic of the optical setup used in the experiment. Probe pulses for the polarization rotation measurements were provided by a femtosecond Ti:sapphire amplifier system and focused onto the sample with a lens to a ~30 µm spot. A narrow bandpass filter (808/5 nm) was used to select a central wavelength of 808 nm. The probe beam was aligned at normal incidence to the sample surface, and its polarization was controlled using a quarter waveplate and a polarizer. The strain-induced polarization rotation was measured using a combination of a half waveplate, a Wollaston prism and two photodiodes. Uniaxial strain was applied to the sample using a strain cell (Razorbill Instruments UC220T). All the measurements were carried out under ambient conditions.

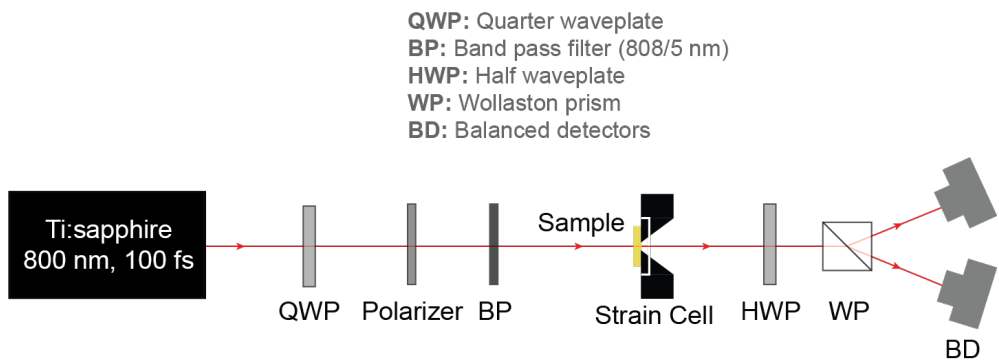

**Figure S1. | Schematic of the experimental setup.**

### S1.2. Sample Preparation

The sample investigated was a commercially available single crystal $AgGaS_2$ of 200 µm thickness with an optically flat (001) surface. Strips of 2.5 mm * 0.5 mm were cut from the same crystal along the crystal *a*- and *b*-axes, respectively, enabling to apply uniaxial strain along both the $x$ and $y$ directions. These strips were mounted onto separate adapters for the strain cell, enabling exchange of samples during the measurement campaign.

## S1.3. Application of Uniaxial Strain

The strain cell employs piezo-stack actuators to generate a tunable uniaxial stress, both compressive and tensile. It is equipped with a sensor that monitors the force applied to the sample. The applied stress $\sigma$ is evaluated from the ratio between the measured force and the sample's cross-section area. The resulting strain $\varepsilon$ is then calculated from the stress using the stiffness tensor $S$ of the crystal[1]

$$\varepsilon_{ij} = S_{ijkl}\sigma_{kl}$$

For uniaxial stress applied along the $x$ direction ($\sigma_{xx}$), the crystal symmetry ($\bar{4}2m$) ensures that all off-diagonal shear strain components are zero. The nonzero strain components are expressed as follows, with the corresponding stiffness constants $S_i$ listed in Table S1.

$$\varepsilon_{xx} = S_1\sigma_{xx}$$
$$\varepsilon_{yy} = S_2\sigma_{xx}$$
$$\varepsilon_{zz} = S_3\sigma_{xx}$$

For uniaxial stress applied along the orthogonal $y$ direction ($\sigma_{yy}$), the relations are

$$\varepsilon_{yy} = S_1\sigma_{yy}$$
$$\varepsilon_{xx} = S_2\sigma_{yy}$$
$$\varepsilon_{zz} = S_3\sigma_{yy}$$

| Stiffness Constant | Value ($10^{-12}$ Pa$^{-1}$) |
|---|---|
| $S_1$ | 26.2 |
| $S_2$ | -7.7 |
| $S_3$ | -14.5 |

Table S1. | Relevant stiffness constants of AgGaS$_2$[1].

## S2. Analysis of the Polarization Rotation Signal

In the experiments, the probe pulses propagate along the optical axis (c-axis) of the crystal, with polarization in the a-b plane. Hence, only the in-plane permittivity tensor elements are relevant and considered as follows.

For unstrained AgGaS$_2$ with $\bar{4}2m$ (D2d) symmetry, the in-plane permittivity is isotropic:

$$\boldsymbol{\varepsilon} = \begin{pmatrix} \varepsilon_{11} & 0 \\ 0 & \varepsilon_{22} \end{pmatrix} = \begin{pmatrix} \varepsilon_{11} & 0 \\ 0 & \varepsilon_{11} \end{pmatrix}.$$

Uniaxial strain along the crystal *a* or *b* axis reduces the symmetry of the system to a chiral point group 222 (D2), inducing both birefringence and optical activity. The birefringence modifies the diagonal elements of the permittivity tensor

$$\varepsilon' = \begin{pmatrix} \varepsilon_{11} - b & 0 \\ 0 & \varepsilon_{11} + b \end{pmatrix},$$

while the strain-induced optical activity ($g_{33}$) introduces antisymmetric imaginary off-diagonal terms following

$$\Delta\varepsilon_{ij} = i\, l_{ijk} g_{kl} k_l$$

where $l_{ijk}$ is the levi-civita symbol and $k$ is the wavevector.

Combining both effects, the permittivity tensor is given by

$$\varepsilon'' = \begin{pmatrix} \varepsilon_{11} - b & ic \\ -ic & \varepsilon_{11} + b \end{pmatrix},$$

with

$$c = g_{33} k_3$$

the effective optical activity and $b$ the effective birefringence.

We note that for unstrained AgGaS$_2$ with $\bar{4}2m$ (D2d) symmetry, the optical activity tensor is diagonal, with $g_{11} = -g_{22}$ and $g_{33} = 0$. Since the probe pulses propagate along the optical axis (*c*-axis) of the crystal, only the $g_{33}$ component contributes to the polarization rotation signal and directly reflects the presence of chirality induced by strain.

The incident linearly polarized probe pulse can be decomposed into two eigenvectors following the permittivity tensor in the strained state described above. Defining $f = \frac{c}{b}$, the eigenvectors in the *a-b* plane are two elliptically polarized light waves with opposite ellipticity.

$$\boldsymbol{e_1} = \begin{pmatrix} \dfrac{i(-1+\sqrt{1+f^2})}{f\sqrt{1+\dfrac{(-1+\sqrt{1+f^2})^2}{f^2}}} \\[2ex] \dfrac{1}{\sqrt{1+\dfrac{(-1+\sqrt{1+f^2})^2}{f^2}}} \end{pmatrix} \qquad \boldsymbol{e_2} = \begin{pmatrix} \dfrac{-i(1+\sqrt{1+f^2})}{f\sqrt{1+\dfrac{(1+\sqrt{1+f^2})^2}{f^2}}} \\[2ex] \dfrac{1}{\sqrt{1+\dfrac{(1+\sqrt{1+f^2})^2}{f^2}}} \end{pmatrix}$$

The effective refractive indices for the two eigenvectors are

$$n_1 = \sqrt{\varepsilon_{11} + \sqrt{b^2 + c^2}} \approx \sqrt{\varepsilon_{11}} + \frac{\sqrt{b^2 + c^2}}{2\sqrt{\varepsilon_{11}}}$$

$$n_2 = \sqrt{\varepsilon_{11} - \sqrt{b^2 + c^2}} \approx \sqrt{\varepsilon_{11}} - \frac{\sqrt{b^2 + c^2}}{2\sqrt{\varepsilon_{11}}}.$$

These approximations hold under the conditions $|b| \ll \varepsilon_{11}$ and $|c| \ll \varepsilon_{11}$.

The phase delay $\gamma$ between the two eigenvectors is the product of the sample thickness $d$ and the difference in refractive index.

$$\gamma = \frac{2\pi d}{\lambda_0}(n_1 - n_2) \approx \frac{2\pi d}{\lambda_0}\frac{\sqrt{b^2 + c^2}}{\sqrt{\varepsilon_{11}}}$$

The two eigenvectors are then recombined after passing through the sample. The polarization rotation signal is analyzed with a half waveplate and Wollaston prism.

Using the Jones matrix analysis[2], the effective polarization rotation $\theta$ is proportional to the normalized intensity difference on the two detectors following

$$\theta(\varphi) = \frac{I_1 - I_2}{2(I_1 + I_2)} = \frac{1}{2\sqrt{1 + 1/f^2}}\sin(\gamma) + \frac{(\cos(\gamma) - 1)}{4(1 + f^2)}\sin(4\varphi).$$

Here, $\varphi$ is the incident polarization angle relative to the strain axis. Expanding $\sin(\gamma) \approx \gamma$ and $\cos(\gamma) \approx 1 - \frac{\gamma^2}{2}$, we arrived at

$$\theta(\varphi) = \frac{\pi d}{\lambda_0 \sqrt{\varepsilon_{11}}}c - \frac{\pi^2 d^2}{2\lambda_0^2 \varepsilon_{11}}\sin(4\varphi)b^2.$$

It is clear from this expression that the signal is composed of two contributions. The first one is proportional to the optical activity coefficient $c$ and does not depend on the incident probe polarization. The second contribution is proportional to the square of the birefringence coefficient, $b^2$, and depends on the incident probe polarization $\varphi$ with a 90°-periodicity, consistent with the experimental observations. Importantly, the birefringence contribution averages to zero when integrated over all incident probe polarizations allowing us to extract the optical activity from measurements of the polarization rotation for all the 360° incident probe polarization angles.

The strain-induced change in the permittivity tensor (Fig. S2) can thus be acquired based on the experimental results and the parameters listed in Table S2. Both the effective birefringence $b$ and optical activity $c$ are much smaller than $\varepsilon_{11}$, validating the approximations used in the analysis above.

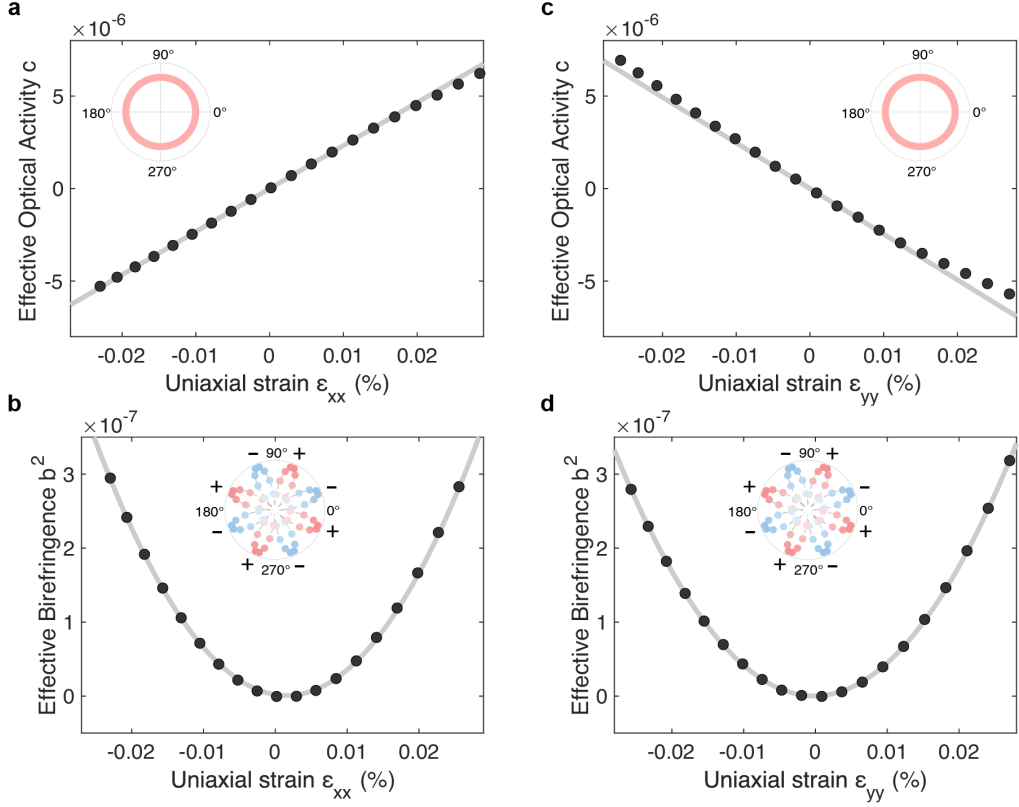

**Figure S2 | Strain-induced changes in permittivity.** (a) Effective optical activity coefficient $c$ as a function of uniaxial strain along the $x$ direction. (b) Square of the effective birefringence coefficient $b^2$ as a function of uniaxial strain along the $x$ direction. (c) Same as panel (a) as a function of uniaxial strain along the $y$ direction. (d) Same as panel (b) as a function of uniaxial strain along the $y$ direction.

| Physical Quantity | Value |
|---|---|
| Sample thickness $d$ | 200 µm |
| Probe wavelength $\lambda_0$ | 808 nm |
| Static permittivity[3] $\varepsilon_{11}$ | 6.19 |

**Table S2 | Relevant experimental parameters.**

Finally, the rotary power $\rho$ is obtained as

$$\rho = \frac{\pi c}{\lambda_0 \sqrt{\varepsilon_{11}}}.$$

Figure S3 provides an intuitive illustration of the birefringence contribution to the polarization rotation signal. When a linearly polarized pulse passes through a birefringent crystal, the transmitted beam becomes elliptically polarized. The projection on the $x$- and $y$-axes (the balancing axes for the incoming polarization) will differ depending on the relative angle between the incoming polarization and the fast axis of the birefringent material. As a result, the signal exhibits a four-fold symmetry with respect to the incoming polarization.

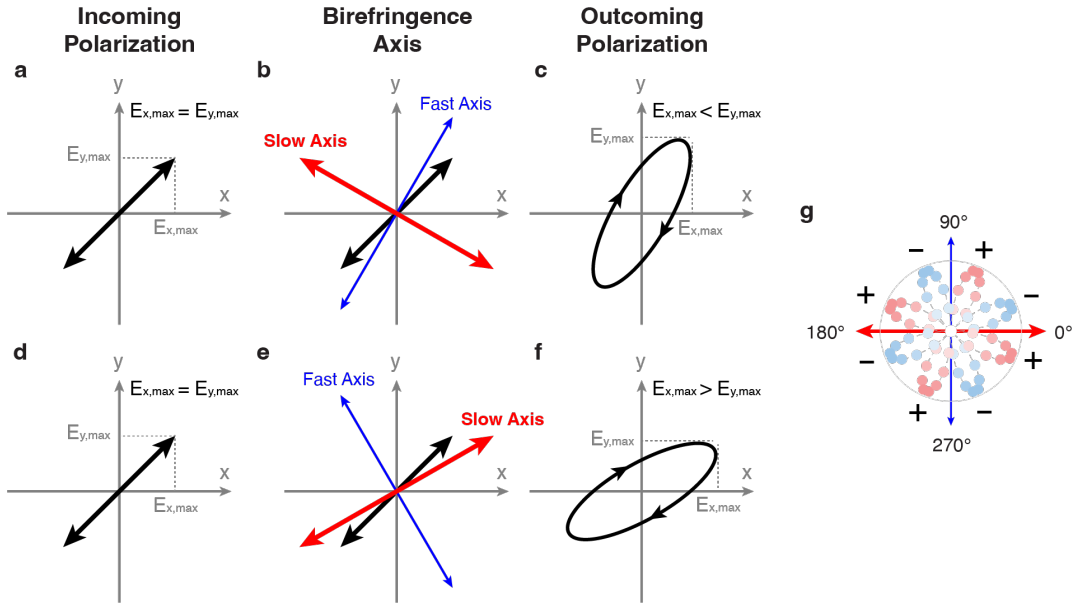

**Figure S3 | Illustration of the contribution of birefringence to the polarization rotation signal. (a) Linearly polarized incoming polarization. X and Y axes are set to balance the signal on the two detectors. (b) Orientation of the birefringence axis relative to the incoming polarization with fast axis closer to the incoming polarization (c) The outcoming polarization, where the projections on the two detectors differ. (d)-(f) Orientation of the slow axis closer to the incoming polarization. The sign of the signal is inversed. (g) Polarization rotation signal as a function of the incident polarization angle.**

## S3. Symmetry Analysis of Piezochiral Tensor and Higher-order Piezochiral Tensor

Since strain does not break inversion symmetry, the piezochiral effect can only occur in achiral systems without inversion symmetry (10 crystallographic point groups).

From a symmetry perspective, chirality is a pseudoscalar quantity with Jahn symbol[4,5] (e) while strain is a 2$^{nd}$ rank symmetric tensor, with Jahn symbol ($[V^2]$). Therefore, their linear coupling corresponds to a 2$^{nd}$ rank pseudotensor ($e[V^2]$), which is parity odd and time-reversal even.

Among all the achiral point groups, only four allow such linear strain-chirality coupling.

For the higher-order piezochiral chiral tensor, the general coupling term for the n$^{th}$ order writes

$$C = \sum_{i_1 j_1 i_2 j_2 \ldots i_n j_n} T_{i_1 j_1 i_2 j_2 \ldots i_n j_n} \varepsilon_{i_1 j_1} \varepsilon_{i_2 j_2} \ldots \varepsilon_{i_n j_n}$$

where the coupling tensor transforms with the symmetry of $e[[V^2]^n]$.

There are four groups with the 2nd order piezochiral coupling as the leading term, with the following coupling forms.

| Point group symmetry | Piezochiral coupling |
|---|---|
| 4mm (C4v) | $C = \theta(\varepsilon_{xx} - \varepsilon_{yy})\varepsilon_{xy}$ |
| 3m (C3v) | $C = \theta[(\varepsilon_{xx} - \varepsilon_{yy})\varepsilon_{xz} - 2\varepsilon_{xy}\varepsilon_{yz}]$ |
| -6m2 (D3h) | $C = \theta[(\varepsilon_{xx} - \varepsilon_{yy})\varepsilon_{xz} - 2\varepsilon_{xy}\varepsilon_{yz}]$ |
| -6 (C3h) | $C = \theta_1[(\varepsilon_{xx} - \varepsilon_{yy})\varepsilon_{xz} - 2\varepsilon_{xy}\varepsilon_{yz}]$ $+ \theta_2[(\varepsilon_{xx} - \varepsilon_{yy})\varepsilon_{yz} + 2\varepsilon_{xy}\varepsilon_{xz}]$ |

**Table S3 | Second-order piezochiral effect. Point group symmetries and their corresponding piezochiral coupling forms.**

There are two groups with the 3rd order piezochiral coupling as the leading term

| Point group symmetry | Piezochiral coupling |
|---|---|
| -43m (Td) | $C = \theta_1(\varepsilon_{xx} - \varepsilon_{yy})(\varepsilon_{yy} - \varepsilon_{zz})(\varepsilon_{zz} - \varepsilon_{xx})$ $+ \theta_2[(\varepsilon_{xx} - \varepsilon_{yy})\varepsilon_{xy}^2 + (\varepsilon_{yy} - \varepsilon_{zz})\varepsilon_{yz}^2 + (\varepsilon_{zz} - \varepsilon_{xx})\varepsilon_{xz}^2]$ |
| 6mm (C6v) | $C = \theta_1[\varepsilon_{xy}(\varepsilon_{yz}^2 - \varepsilon_{xz}^2) + \varepsilon_{xz}\varepsilon_{yz}(\varepsilon_{xx} - \varepsilon_{yy})]$ $+ \theta_2[\varepsilon_{xy}(3\varepsilon_{xx}^2 + 3\varepsilon_{yy}^2 - 6\varepsilon_{xx}\varepsilon_{yy} - 4\varepsilon_{xy}^2)]$ |

**Table S4 | Third-order piezochiral effect. Point group symmetries and their corresponding piezochiral coupling forms.**

In total, all 10 achiral non-centrosymmetric point groups allow piezochiral coupling: four with first order, four with second order, and two with third order contributions. The systematic classification is summarized in Figure S4.

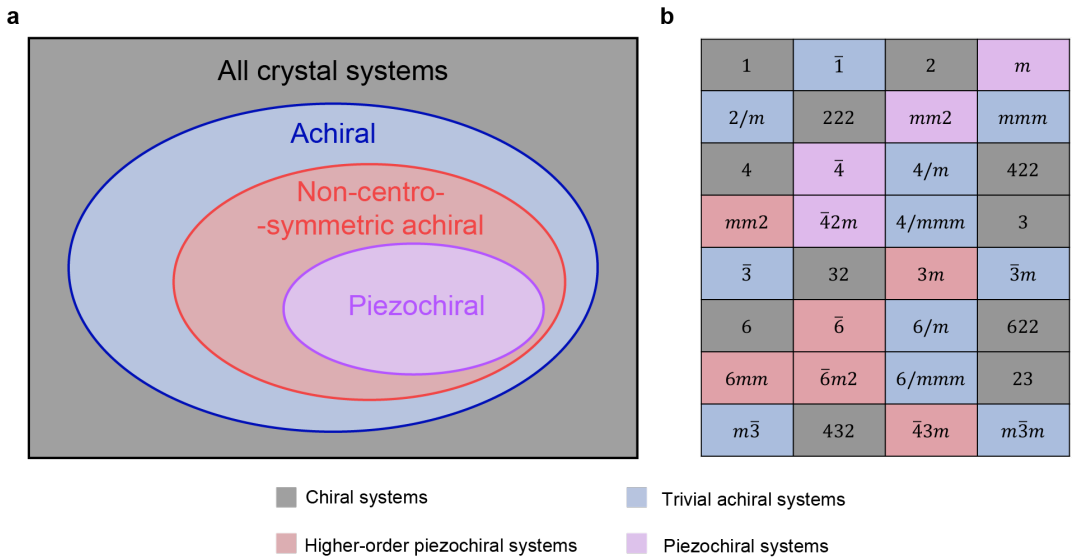

**Figure S4. | Classification of crystal systems. (a)** Piezochiral systems form a subset of non-centrosymmetric achiral systems. **(b)** Classification of 32 crystallographic point groups: 11 groups (gray) are chiral, 11 groups (blue) are trivial achiral, 6 groups (red) permit higher-order piezochirality, and 4 groups (purple) are piezochiral.

In practice, these higher-order effects can be accessed by applying specific combinations of strain components. For instance, tensile strain applied along the [111] body diagonal in cubic GaAs (point group $\bar{4}3m$) generates equal strain in $\varepsilon_{xx}$, $\varepsilon_{yy}$ and $\varepsilon_{zz}$, which, according to the symmetry-allowed coupling form, leads to the emergence of chirality.

## S4. Chiral Order Parameter Calculations

We performed first-principles calculations in the framework of density functional theory (DFT) to explore the impact of strain on the structure of AgGaS$_2$. Our computations are carried out with the Vienna *Ab initio* Simulation Package (VASP, version 6.5.1)[6-8], using pseudopotentials generated within the Projector Augmented Wave (PAW) method[9]. To parametrize the exchange-correlation energy, we applied the PBEsol[10] approximation and used the default potentials of the VASP package for Ga, Ag, and S.

For the numerical settings, we sampled the reciprocal space with an 8×8×4 k-point mesh generated using the scheme of Monkhorst and Pack[11]. We also set the plane-wave cutoff to 550 eV. Both settings were applied consistently to all our computations. Additionally, for all calculations, we iterated the self-consistent cycle until the change in total energy was smaller than $10^{-8}$ eV.

The starting point of our investigation is the structural relaxation of the AgGaS$_2$ unit cell. Within PBEsol, we find the equilibrium volume to be 339.43 Å$^3$ (a = b = 5.66 Å, c = 10.55 Å). The atomic equilibrium positions were obtained at the following Wyckoff sites: Ga (4a), Ag (4b), and S (8d, x = 0.77944). Using this relaxed structure, we performed additional structural relaxations under the constraint of a fixed, modulated a- or b-axis. This modulation breaks the tetragonal symmetry, as discussed above, and induces a net chiral state. The resulting atomic positions were then used to compute the chiral order parameter according to Eq. (4) and as shown in Fig. 3 of the main manuscript.